\documentstyle[twoside,fleqn,epsf,epsfig,espcrc2]{article}
\newskip\humongous \humongous=0pt plus 1000pt minus 1000pt
\def\caja{\mathsurround=0pt}
\def\eqalign#1{\,\vcenter{\openup1\jot \caja
        \ialign{\strut \hfil$\displaystyle{##}$&$
        \displaystyle{{}##}$\hfil\crcr#1\crcr}}\,}
\newif\ifdtup

\def\eqright #1\cr{\noalign{\hfill$\displaystyle{{}#1}$}}
\def\eqleft #1\cr{\noalign{\noindent$\displaystyle{{}#1}$\hfill}}

\def\oldreffmt#1{\rlap{[#1]} \hbox to 2\parindent{}}

\def\figfmt#1{\rlap{Figure {#1}} \hbox to 1in{}}

\def\auto{\eqno(\refstepcounter{equation}\theequation)}
\def\begineq #1\endeq{$$ \refstepcounter{equation}\eqalign{#1}\eqno
	(\theequation) $$}
\def\contlimit{\,{\hbox{$\longrightarrow$}\kern-1.8em\lower1ex
\hbox{${\scriptstyle (a\rightarrow0)}$}}\,}
\def\centeron#1#2{{\setbox0=\hbox{#1}\setbox1=\hbox{#2}\ifdim
\wd1>\wd0\kern.5\wd1\kern-.5\wd0\fi
\copy0\kern-.5\wd0\kern-.5\wd1\copy1\ifdim\wd0>\wd1
\kern.5\wd0\kern-.5\wd1\fi}}
\def\centerover#1#2{\centeron{#1}{\setbox0=\hbox{#1}\setbox
1=\hbox{#2}\raise\ht0\hbox{\raise\dp1\hbox{\copy1}}}}
\def\centerunder#1#2{\centeron{#1}{\setbox0=\hbox{#1}\setbox
1=\hbox{#2}\lower\dp0\hbox{\lower\ht1\hbox{\copy1}}}}
\def\lsim{\;\centeron{\raise.35ex\hbox{$<$}}{\lower.65ex\hbox
{$\sim$}}\;}
\def\gsim{\;\centeron{\raise.35ex\hbox{$>$}}{\lower.65ex\hbox
{$\sim$}}\;}
\def\st#1{\centeron{$#1$}{$/$}}

\def\super#1{\ifmmode \hbox{\textsuper{#1}}\else\textsuper{#1}\fi}
\def\textsuper#1{\newcount\holdspacefactor\holdspacefactor=\spacefactor
$^{#1}$\spacefactor=\holdspacefactor}

\def\getcite#1,{\advance\citenumber by1
\ifnum\citenumber=1
\ref{#1}\let\next=\getcite\else\ifx#1@\let\next=\relax
\else ,\ref{#1}\let\next=\getcite\fi\fi\next}

\def\upon #1/#2 {{\textstyle{#1\over #2}}}
\relax

\def\mainhead#1{\setcounter{equation}{0}\addtocounter{section}{1}
  \vbox{\begin{center}\large\bf #1\end{center}}\nobreak\par}

\def\til#1{\centeron{\hbox{$#1$}}{\lower 2ex\hbox{$\char'176$}}}
\def\tild#1{\centeron{\hbox{$\,#1$}}{\lower 2.5ex\hbox{$\char'176$}}}
\def\sumtil{\centeron{\hbox{$\displaystyle\sum$}}{\lower
-1.5ex\hbox{$\widetilde{\phantom{xx}}$}}}

\def\kbar{\underline{k}}

\def\pom{{\rm P\kern -0.53em\llap I\,}}
\def\spom{{\rm P\kern -0.36em\llap \small I\,}}
\def\sspom{{\rm P\kern -0.33em\llap \footnotesize I\,}}

\newcommand{\AmS}{{\protect\the\textfont2
  A\kern-.1667em\lower.5ex\hbox{M}\kern-.125emS}}

\title{\rightline{ANL-HEP-CP-00-92}
\rightline{\today} 
$~$\\
$~$\\
$~$\\
$~$\\
\centerline{THE ANOMALY AND REGGEON FIELD THEORY IN QCD$^*$}
$~$\\
$~$
}

\author{\centerline{ALAN R. WHITE$^{**}$} 
$~$\\
\centerline{High Energy Physics Division,} 
\centerline{Argonne National Laboratory,}
\centerline{Argonne, Il 60439, U.S.A.}
} 
       
\begin{document}

\begin{abstract}
$~$

\vspace{0.5in}

$~$
The appearance of the U(1) anomaly in the interactions of reggeized gluons
is described. Also discussed is the crucial role the anomaly can play 
in providing 
the non-perturbative properties necessary for a transition
from gluon and quark reggeon diagrams to hadron reggeons and a 
reggeon field theory description of the pomeron. 

\vspace{1.5in}

\centerline{Presented at the International Euroconference 
on Quantum Chromodynamics QCD 00,}
$~$\\
\centerline{July 6-13, Montpellier, France.}

\vspace{0.7in}

\noindent $^*$ Work
supported by the U.S.
Department of Energy, Division of High Energy Physics, Contracts
W-31-109-ENG-38 and DEFG05-86-ER-40272

\noindent $^{**}$ arw@hep.anl.gov 

\end{abstract}

\maketitle

\section{INTRODUCTION}

In this talk I will report on a program\cite{arw98,arw99}, the 
goal of which is, to find a (candidate) regge region QCD S-Matrix 
starting from  reggeized gluons and quarks (reggeons).
We look for hadrons and the 
pomeron as bound-states of quark and gluon reggeons that appear 
in multi-regge amplitudes. The dynamics involved will be the infra-red 
divergences that occur in reggeon diagrams when 
both gluons and quarks are massless.

To obtain a unitary pomeron, we anticipate that Reggeon Field Theory (RFT), 
and in particular the critical pomeron\cite{cri},
will be essential. RFT describes the pomeron as a regge pole plus
multipomeron exchanges and interactions,  
in agreement with experiment, but not with perturbative QCD. 
The critical pomeron is the only known solution of $s$-
and $t$-channel unitarity that gives a rising total cross-section 
and a connection with QCD could have important consequences. Firstly, since it
allows physical cross-sections to have scaling behavior (up to logarithms), a
rising total cross-section should correspond to the maximal applicability of
asymptotic freedom to short distance processes. In addition, the factorization
properties of the critical pomeron provide
a ``universal wee-parton distribution'' that could carry the (vacuum)
properties of confinement and chiral-symmetry breaking as part of
an extension of the parton model beyond leading-twist perturbation theory. 

Since our reggeon diagram starting point is essentially 
perturbative it might be expected that 
the ``non-perturbative'' properties
of confinement and chiral-symmetry breaking will not appear at all in our 
formalism. In this talk, however, we will show that 
the U(1) anomaly appears in particular interactions of 
gluon reggeons\cite{arw99}. The anomaly 
represents the potential ultra-violet/infra-red flow of chirality into and
out of the theory and we will indicate how, when appropriately treated, 
it gives an infra-red divergence that can 
combine with the infra-red divergences of gluon reggeon 
diagrams to produce a transition to hadron and pomeron reggeon diagrams 
in which the desired non-perturbative
properties appear (as properties of the spectrum).

\section{REGGEON DIAGRAMS FROM SPONTANEOUSLY-BROKEN QCD}

We begin by summarizing well-known results\cite{fkl}
from perturbative calculations
in spontaneously-broken gauge theories. 
The regge limit of an elastic scattering amplitude, i.e. 
$s \to \infty$, $t$ fixed, is studied in
the complex angular momentum $J $-plane via the (Sommerfeld-Watson) 
representation 
$$
A(s,t) ~= ~ \int ~d J ~a(J,t)~s^J
\auto
$$
When all gluons and quarks
have a mass there are no infra-red divergences and the regge behavior
is straightforward.
Leading-log 
calculations show that both gluons and quarks ``reggeize''. In 
particular, the gluon becomes a regge pole with trajectory $J=1 - \Delta(t)$.  
Non-leading logs are reproduced by ``reggeon diagrams'', which are 
$k_{\perp}$ integrals  
with gluon particle poles in addition to reggeon propagators. For example,
the two gluon reggeon state appears in $a(J,t)$ as 
$$
\eqalign{&\int {d^2k_1 \over k_1^2 +M^2}~{ d^2k_2  \over k_2^2 + M^2}~
{\delta^2(Q -k_1-k_2)
\over J-1 +  \Delta(k_1^2) + \Delta(k_2^2)} \cr
&~}
\auto
$$

Reggeon unitarity\cite{arw98,gpt,arw1} requires that a complete set of reggeon 
diagrams, involving all possible $J$-plane multi-reggeon states, appear in 
higher-orders. The well-known
BFKL equation is a simple consequence of 2-reggeon unitarity,
i.e. if we denote the 2-reggeon state 
by $\raisebox{-2mm}{$\epsfxsize=0.3in 
\epsffile{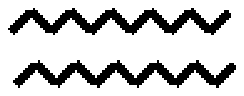}$}$ this equation sums the set of diagrams 
\begin{center}
\epsfxsize=3in 
\epsffile{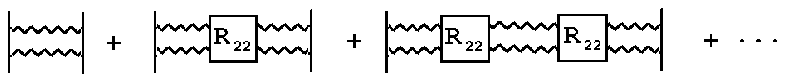}
\end{center}
obtained by iterating 
the $2-2$ reggeon interaction   
$$
\eqalign{&R_{22}=
[(\kbar^2_1+M^2)({\kbar^2_2}'+M^2) \cr
&+(\kbar^2_2+M^2)(
{\kbar^2_1}'+M^2)]/[ (\kbar_1-\kbar_1')^2+M^2] \cr 
&~+ ~\cdots }
\auto
$$

When $M \to 0$, infra-red divergences
exponentiate to zero all diagrams with non-zero $t$-channel color.
Since the gluon poles remain, however, the color zero amplitudes
scale canonically ($\sim k_{\perp}^{-2}$)~, i.e. there is
no confinement! Our aim is to see the gluon poles disappear via  
an RFT phase-transition from quark and gluon to hadron
and pomeron reggeon diagrams.

In general, reggeon unitarity requires\cite{arw98,arw1} that
quark and gluon reggeon diagrams
describe not only elastic scattering but also 
all multi-regge limits of multiparticle amplitudes.
Consequently we can study limits of high-order amplitudes,
an example of which is
illustrated in Fig.~1, where   
both hadrons and the pomeron could appear as
(coupled) bound states of quark and gluon reggeons.
Since we expect the anomaly to be involved and we consider it's
infra-red manifestation, we look for such bound states 
when both the gluon mass $M$ and the quark mass 
$m ~ \to 0$. As we will see, both 
the U(1) anomaly and RFT phase-transition analysis 
are essential for 
understanding the infra-red divergence structure we find. 
To provide 
motivation and to describe features that we will be 
\begin{center}
\leavevmode
\epsfxsize=2.7in
\epsffile{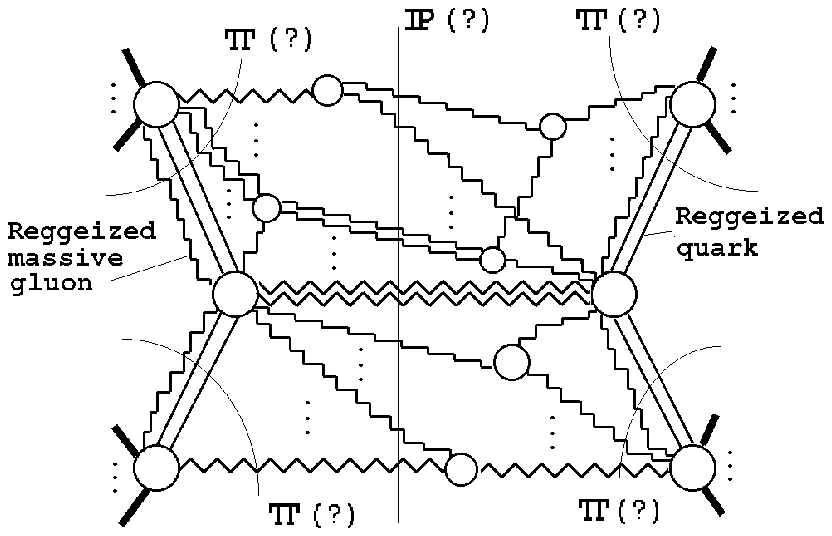}

\vspace{0.2in}

\leavevmode
\epsfxsize=2.7in
\epsffile{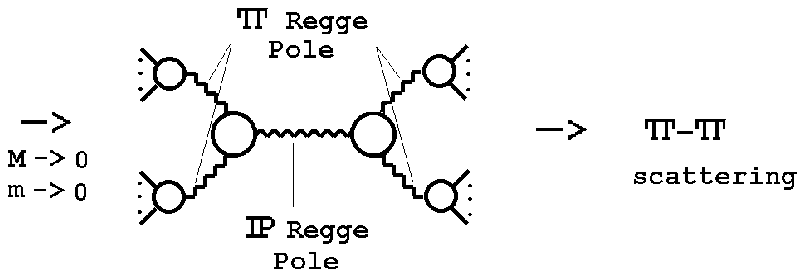}

Fig.~1 An amplitude in which both pion and pomeron regge poles can appear.
\end{center}
looking for,
we first introduce the RFT critical pomeron and an associated 
supercritical phase.

\section{POMERON RFT} 

For a regge pole pomeron, RFT can be 
formulated directly from reggeon unitarity.
The diagrams are essentially the
same\cite{arw98} as for reggeized gluons, but without the gluon poles!

\subsection{The Critical Pomeron}

This is a renormalization group fixed-point 
solution\cite{cri} 
of RFT in which $\Delta (0) = 1 - \alpha_{\spom}(0) = 0 $ and 
total cross-sections rise asymptotically 
$$ 
\sigma_T\centerunder{$\large\sim$}{\raisebox{-4mm}
{$\scriptstyle s\to \infty$}}\;
[\ln s]^\eta
\auto
$$
with $\eta={\epsilon\over 12} +
O(\epsilon^2)$, where $4 - \epsilon$ is the 
transverse momentum dimension. 
Also, all other asymptotic predictions satisfy unitarity  
in both the $t$-channel and the $s$-channel. 

\subsection{The Supercritical Pomeron}

To find the new phase that appears at the critical point, we consider
the critical lagrangian
$$
{\cal L}={1\over 2}\bar\phi
\centerover{${\partial\over \partial y}$}{$\leftrightarrow$}
\phi-\alpha'_0\nabla\bar\phi\nabla\phi - \Delta_0\bar\phi\phi -
{1\over 2}ir_0\left[\bar\phi\phi^2 + \bar\phi^2\phi\right]
\auto
$$
where $\bar\phi$ ($\phi$) creates (destroys) pomerons.
The stationary point
$$
\phi = \bar\phi = {2i\Delta_0\over 3r_0}
\auto
$$ 
gives a pomeron condensate that shifts the pomeron intercept ($ \Delta_0
\to - \Delta_0 /3$) and produces 
a solution\cite{arw1} for $\Delta_0 < 0~$        
($\alpha_{\spom}(0) > 1$). To determine the resulting graphical expansion 
the pomeron condensate has to be be interpreted as a ``wee-parton''
component of the scattering states\cite{arw1}. 
(This solution was very controversial 
20 years ago - although it was supported by Gribov !)

The condensate generates new classes of RFT diagrams, a simple example of 
which is shown in Fig.~2. 
\begin{center}
\leavevmode
\epsfxsize=2.8in
\epsffile{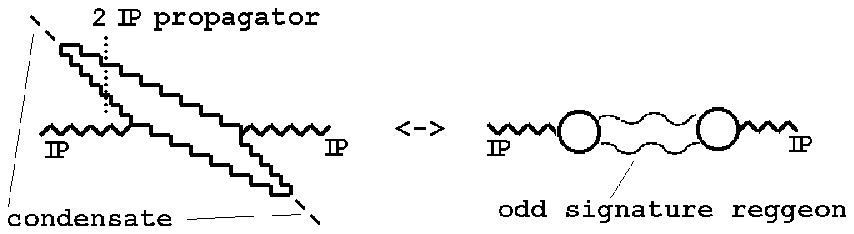}

\vspace{0.1in}

Fig.~2 A new RFT diagram generated by the pomeron condensate

\end{center}
As illustrated in this diagram, 
the two pomeron propagators produced by the condensate  
(i.e. $[ 2 \Delta_0 - 2\alpha_0'~ 
k_{\perp}^2 + \cdots]^{-1}$)  
give $k_{\perp}$ poles that have to be interpreted as
particle poles, implying that there is a pomeron
transition to a two vector reggeon state as shown.
Reggeon states involving many vector particle poles  
similarly appear in higher-order diagrams.

In general, divergences in rapidity produced in the original
graphical expansion because $\Delta_0 < 0$  are 
converted to vector particle divergences in $k_{\perp}$ in the supercritical
expansion. Therefore, the supercritical phase is characterized by the 
``deconfinement of a vector particle on the pomeron trajectory''.

\subsection{The Supercritical Phase and Color Superconductivity}

An immediate question is whether the gluon 
particle poles in QCD could disappear via the reverse 
of the deconfinement process just described. The appearance of a 
reggeized vector particle suggests that
the supercritical phase could be realized in QCD
via the breaking of the gauge symmetry from SU(3) to SU(2). In current 
terminology the supercritical pomeron would correspond to
``color superconducting QCD''.
Identification of the vector mass with the RFT order parameter
would then determine that the critical pomeron appears as the gauge symmetry
is restored. Therefore, to look for 
the critical pomeron we begin by studying
superconducting QCD.

The breaking of SU(3) gauge symmetry to SU(2) produces 
an (odd signature) SU(2) singlet, massive, reggeized gluon. An 
even signature pomeron would be produced if this massive reggeon appears
in an SU(2) singlet reggeon condensate containing an odd number
($\geq 3$) of massless gluon reggeons.
The corresponding SU(3) pomeron would contain an even number ($\geq 4$) of 
gluons but carry odd (``anomalous'')
color parity, in contrast to the even color parity 
($\geq 2$ gluon) BFKL pomeron!

A three gluon condensate carries the quantum numbers of the
winding-number current and so could be due to  
spectral flow of the Dirac sea produced by the anomaly.
If we can understand the origin of a condensate of this kind then, provided
the identification with the supercritical pomeron can be made, the 
disappearance of the condensate will give the critical
pomeron we are looking for.
For this purpose, we must determine 
how the anomaly appears in
the regge-limit effective theory described by
gluon and quark reggeon diagrams? (It is,
of course, absent in the usual perturbation expansion of a vector theory.)
The rest of this talk will be primarily devoted to this issue.
The consequences (RFT for the pomeron, confinement and chiral symmetry
breaking etc.) will be outlined only briefly.

\section{THE ANOMALY IN TRIPLE-REGGE VERTICES}

The simplest multi-regge limit in which the anomaly makes an appearance is the 
full triple-regge limit\cite{gw}. It is present in the ``helicity-flip''
part\cite{arw99} of reggeized gluon interactions containing a single 
quark loop. (That bound-state couplings involve helicity-flip effects
is directly related to the occurrence of chiral symmetry breaking.)

Consider the 3-3 scattering of quarks, each of which has a large 
light-cone momentum, but with the spacelike components of the momenta 
orthogonal, i.e.  
$$
\eqalign{ P_1~\to&~ P_1^+~= ~(p_1,p_1,0,0)~,~~p_1 \to \infty \cr
P_2~\to&~ P_2^+~= ~(p_2,0,p_2,0)~,~~p_2 \to \infty \cr
P_3~\to&~ P_3^+~= ~(p_3,0,0,p_3)~,~~p_3 \to \infty  }
\auto
$$
with the $Q_i$ finite. We consider diagrams of the form illustrated in
Fig.~3.
\begin{center}
\leavevmode
\epsfxsize=1.6in
\epsffile{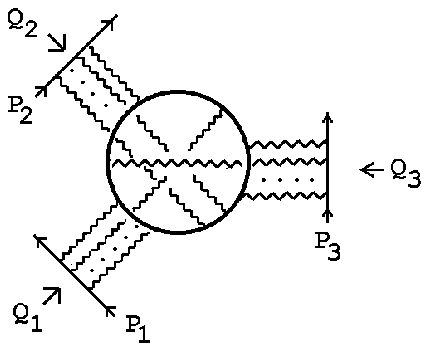}

Fig.~3 Diagrams containing a single quark loop.
 
\end{center}
Using light-cone co-ordinates, the leading behaviour is 
obtained by putting quark lines on-shell via $k_{i^{\pm}}$ integrations
leaving the $k_{\perp}$-integrals of 
reggeon diagrams,
as illustrated in Fig.~4.
\begin{center}
\leavevmode
\epsfxsize=2.7in
\epsffile{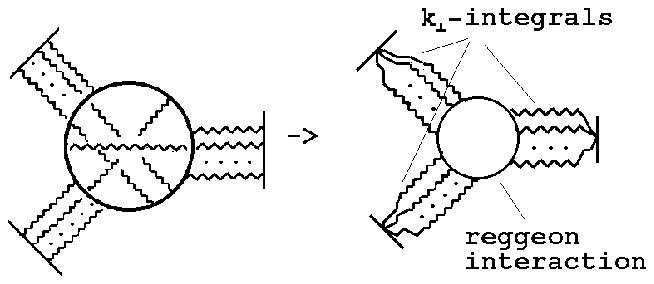}

Fig.~4 The reduction to reggeon diagrams.

\end{center}
The resulting ``triple-regge'' reggeon interactions are quark
triangle diagrams with both local (point-like) and non-local
``effective vertices'' containing $\gamma$-matrix products. In some
some cases $\gamma_5 \gamma$ 
couplings appear that, potentially, could generate the
triangle anomaly.

A particular example, which we consider in more detail below, is the 
``maximally non-planar'' diagram shown in Fig.~5. It is well-known that 
non-planar diagrams provide the essential structure 
of regge cut couplings. In a gauge theory other diagrams also contribute but, 
essentially, produce only subtraction effects that cancel large momentum 
divergences of the non-planar contributions. The anomaly could, however, be 
subject to such cancelations. 

\noindent
\parbox{1in}{
\begin{center}
\epsfxsize=0.8in
\epsffile{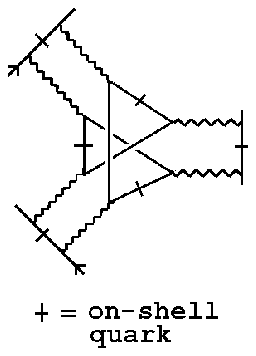}
\end{center}
}
\parbox{1.9in}{
\leavevmode
\epsfxsize=1.7in
\epsffile{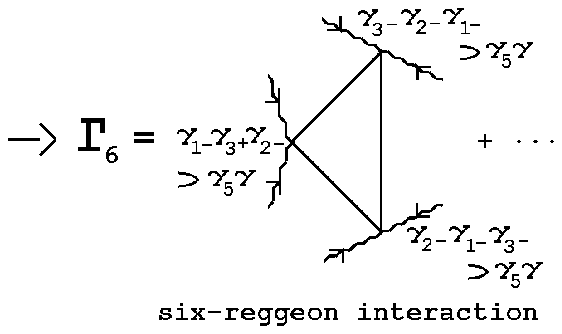}
}
\begin{center}
Fig.~5 The six-reggeon interaction obtained from
a maximally non-planar diagram.
\end{center}

At lowest-order (i.e. two gluons in each $Q_i$ channel) there are 
$\sim 100$ diagrams that potentially could give contributions.
Many obviously do not contain the anomaly while others, in particular 
the maximally non-planar diagrams,
give several contributions. To systematically evaluate all contributions
and also to discuss cancelations a triple-regge 
asymptotic dispersion relation formalism,
in which multiple discontinuities are initially 
calculated rather than amplitudes, can be used\cite{arw99}. A crucial feature 
of this formalism is the contribution, to the dispersion relation,
of unphysical triple discontinuities that contain  
chirality transitions. The absence of the anomaly in 
simpler multi-regge limits can be understood as due to the absence of
such contributions in the corresponding asymptotic dispersion relation.

We will not describe the dispersion relation formalism in this talk.  
Full details can be found in \cite{arw99}.
Instead we will study a maximally non-planar diagram directly.
We will see, however, how the anomaly is associated with an unphysical
multiple discontinuity. First we describe the infra-red properties of the
anomaly that we will look for.

\section{THE ANOMALY AS AN INFRA-RED DIVERGENCE}

To avoid the ultra-violet subtleties of reggeon interactions, and to see a 
connection with spectral flow, we will look for the anomaly 
in the infra-red region\cite{cg}. 
For a massless quark axial-vector current the anomaly 
divergence equation for the three-current vertex gives
\newline\parbox{1.1in}{
\begin{center}
\leavevmode
\epsfxsize=1in
\epsffile{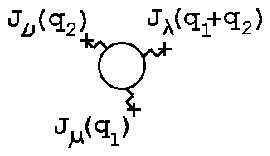}
\end{center}
}
\parbox{1.8in}{ $$  
\sim  ~ \epsilon_{\nu\lambda\alpha\beta} 
{q_{1 \mu}q_2^\alpha (q_1+q_{2})^\beta \over q^2 }
~+ \cdots 
\auto
$$
}
\newline in the limit $ q_1^2 \sim q_2^2 \sim (q_1 +
q_2)^2 \sim q^2~\to 0$.
(For the U(1) current we ignore non-perturbative 
contributions. We are looking for a ``perturbative'' effect!)

If $q_{1^+} \st{\to} 0~$ and $q_2$ is spacelike with $ q_2 \perp q_{1^+}$ 
$$
\epsilon_{\nu\lambda\alpha\beta} ~{q_{1 \mu}q_2^\alpha q_1^\beta 
\over q^2} ~ \sim ~{ {q_{1^+}}^2 \over q} ~\sim ~ {1 \over q} 
\auto
$$
(In the absence of a Lorentz-covariant 
separation of kinematic factors, this linear divergence can be used 
to characterize the anomaly.) 
  
As illustrated in Fig.~6, the divergence is produced by a 
zero momentum chirality transition (corresponding to spectral flow)
combined with a
light-like momentum $q_{1^+}$ flowing through the other two propagators
in the loop.
\begin{center}
\leavevmode
\epsfxsize=2in
\epsffile{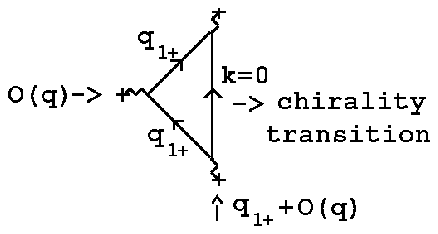}

Fig.~6 The momentum configuration producing the anomaly. 

\end{center}
Because the chirality transition in the quark loop is essential, 
the anomaly infra-red divergence 
can not be canceled by gluon diagrams.

\section{A MAXIMALLY NON-PLANAR DIAGRAM}

In this Section we give an abbreviated version of the calculation, presented
in detail in \cite{arw99}, of the 
triple-regge limit of the maximally non-planar diagram shown in Fig.~5.
We can redraw the diagram and label momenta as in Fig.~7.

\noindent \parbox{1in}{
\leavevmode
\epsfxsize=1in
\epsffile{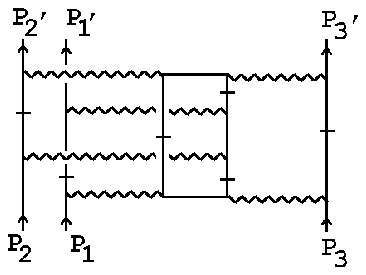}
\newline $~$
}
\parbox{0.2in}{ $$ \to $$
$~$}
\parbox{1.6in}{
\begin{center}
\leavevmode
\epsfxsize=1.6in
\epsffile{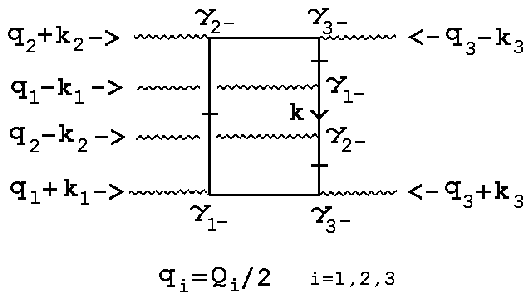}
\end{center}
}
\begin{center}
Fig.~7 Momenta in a maximally non-planar miagram.
\end{center}
For the $k_1$ and $k_2$ integrations 
we use the special light-cone co-ordinates (``generalized Sudakov
variables'') 
$$
k_i=~k_{i2^-}~\underline{n}_{1^+}~+~ 
k_{i1^-}~\underline{n}_{2^+}~+~\underline{\tilde{k}}_{i12}~
~~~i=1,2
$$
where 
$\underline{n}_{1^+} = (1,1,0,0)$ and $\underline{n}_{2^+} = (1,0,1,0)$,
together with the conventional light-cone co-ordinates 
($k_{3^+},k_{3^-},
\underline{\tilde{k}}_{3\perp}$) for the $k_3$ integration. The 
$k_{11^-},k_{22^-}$, $k_{3^-}$ integrations are straightforward.
The six 
options for using the remaining longitudinal $k_i$ integrations to 
put hatched lines on-shell are shown in Fig.~8.
\begin{center}
\leavevmode
\epsfxsize=2.8in
\epsffile{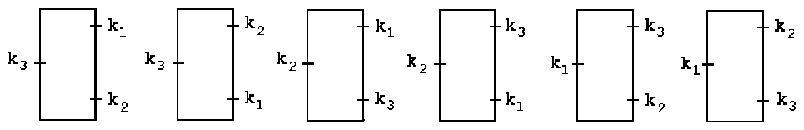}

\vspace{0.1in}

Fig.~8 Options for the longitudinal integrations.
\end{center}

We look for the anomaly to be generated by 
local $\gamma_5 \gamma$ couplings.
A ``local'' coupling is obtained from the component of a
quark numerator with the same 
momentum factor that scales the corresponding
integrated longitudinal momentum.
In the chosen co-ordinates,
only the first option in Fig.~8 gives 
couplings, all three of which have the required components. In this case  
we obtain
$$
\eqalign{ &\int d k_{12^-} ~\delta\biggl( (k_1 + k - q_1)^2 - m^2 \biggr) \cr
&~~~~ \gamma_{3^-} ~\biggl( (k_1 +k - q_1) \cdot 
\gamma + m \biggr) ~ \gamma_{1^-} ~~~~~~ ~~~~~~~~ \cr
&~= ~\int d k_{12^-} ~\delta\biggl( k_{1^-}~ k_{12^-}
 + \cdots \biggr) \cr
&~~~~ \gamma_{3^-} ~
\biggl( k_{1^-}  
\gamma_{2^-} + \cdots \biggr) ~
\gamma_{1^-}\cr
&~ = ~~ \gamma_{3^-} \gamma_{2^-} \gamma_{1^-} ~~ + ~~\cdots }
\auto
$$
$$
\eqalign{&\int d k_{21^-} \delta\biggl( (k_2 - k -q_2)^2 - m^2 \biggr) \cr
&~~~~\gamma_{2^-} ~\biggl( (k_2 -k - q_2 ) \cdot 
\gamma + m \biggr) ~ \gamma_{3^-} ~~~~~~ ~~~~~~~~ \cr
&~= ~\int d k_{21^-} ~\delta\biggl( k_{2^-} k_{21^-}
 + \cdots \biggr) \cr
&~~~~\gamma_{2^-} ~
\biggl( k_{2^-}  
\gamma_{1^-} + \cdots \biggr) ~
\gamma_{3^-} \cr
&~= ~~ \gamma_{2^-} \gamma_{1^-} \gamma_{3^-} ~~ + ~~\cdots }
\auto
$$
$$
\eqalign{&\int d k_{33^+}~ \delta\biggl( (k_3 + k +k_1 -k_2)^2 - m^2 
\biggr) \cr
&~~~~\gamma_{1^-} ~\biggl( (k_3 + k  + k_1 -k_2) \cdot 
\gamma + m \biggr) ~ \gamma_{2^-} \cr
&~=~\int d k_{33^+} ~\delta\biggl( (k_{3^-} + k_{13^-} 
- k_{23^-} )k_{33^+}
 + \cdots \biggr) \cr
& ~~~~ \gamma_{1^-} ~
\biggl( ( k_{3^-} + k_{13^-} 
- k_{23^-} ) 
\gamma_{3^+} + \cdots \biggr) ~
\gamma_{2^-} \cr
&~= ~~ \gamma_{1^-} \gamma_{3^+} \gamma_{2^-} ~~ + ~~\cdots 
}
\auto
$$

Writing
$$
\eqalign{\hat{\gamma}_{31} &=\gamma_{3^-}\gamma_{2^-}\gamma_{1^-}
=\gamma^{-,+,-}- i\gamma_5
\gamma^{-,-,-} \cr
\hat{\gamma}_{23} &=\gamma_{2^-}\gamma_{1^-}\gamma_{3^-}
=\gamma^{+,-,-}- i\gamma_5
\gamma^{-,-,-}  \cr
\hat{\gamma}_{12} &=\gamma_{1^-}\gamma_{3^+}\gamma_{2^-}
=\gamma^{-,-,-}+ i\gamma_5
\gamma^{-,-,+}  }
\auto
$$
where
$$
\eqalign{\gamma^{\pm,\pm,\pm} &= \gamma\cdot n^{\pm,\pm,\pm} \cr
n^{\pm,\pm,\pm} ~&=~ (1,\pm1, \pm1, \pm1) }
\auto
$$
the part of the asymptotic amplitude with local couplings is
$$
\eqalign{& g^{12} {p_{11^+}p_{22^+}p_{3^+} \over m^3} 
 \int { d^2 \underline{k}_{112+} \over
(q_1 + \underline{k}_{112+})^2(q_1 - \underline{k}_{112+})^2} \cr 
&~~~ \int {d^2  \underline{k}_{212} \over
(q_2 + \underline{k}_{212+})^2(q_2 - \underline{k}_{212+})^2} \cr
&~~~\int  {d^2  \underline{k}_{33\perp} \over
(q_3 + \underline{k}_{33\perp})^2(q_3 - \underline{k}_{33\perp})^2} 
\cr  
&\int d^4 k~{ Tr \{ \hat{\gamma}_{12} (\st{k}+ \st{k}_1 
+ \st{q}_2 + \st{k}_3 +m) \hat{\gamma}_{31}
\over 
([k + k_1 + q_2 + k_3]^2 - m^2) } \cr
&~{( \st{k} +m) 
\hat{\gamma}_{23} (\st{k}- \st{k}_2 + \st{q}_1 
+ \st{k}_3 +m) \} \over
(k^2 - m^2)([k - k_2 + q_1 +k_3]^2 - m^2)} }
\auto
$$
For simplicity, we have set the gluon mass to zero. This means that the 
transverse momentum integrals are actually infra-red divergent. However, since
our next step is to
remove both these integrals and the gluon propagators as the lowest-order
contribution of three two-reggeon states, these divergences play no role
in the present discussion. The general structure of gluon transverse momentum
divergences will, of course, play a vital role in the full analysis of reggeon 
diagrams that we outline in the final Section. 

Removing the transverse integrals and the gluon propagators, 
as illustrated in Fig.~5, we obtain a triangle diagram 
six-reggeon vertex $\Gamma_6$. The component with three
$\gamma_5$ couplings is the ($m=0$) triangle 
diagram illustrated in Fig.~9 
\begin{center} 
\epsfxsize=1.8in
\epsffile{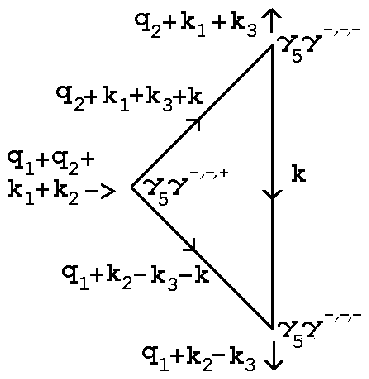}

Fig.~9 The triangle diagram giving the anomaly in $\Gamma_6$.
\end{center}
while the terms with a single $\gamma_5$ coupling have the wrong kinematic 
structure\cite{arw99}. Therefore, we can write
$$
\eqalign{&\Gamma_6(q_1,q_2,q_3,
\tilde{\underline{k}}_1,\tilde{\underline{k}}_2, 
\underline{k}_{3\perp},0) ~~= \cr
\int d^4& k  {  Tr \{ 
\gamma_5 \gamma^{-,-,+} (\st{k}+ \st{k}_1 + \st{q}_2 +\st{k}_3) 
\gamma_5 \gamma^{-,-,-} \over  (k + k_1 + q_2 + k_3 )^2  
~k^2 } \cr
&{\st{k} 
\gamma_5 \gamma^{-,-,-}(\st{k}- \st{k}_2 + \st{q}_1 + \st{k}_3 ) 
\over (k - k_2 + q_1 + k_3)^2 } 
 ~+ ~ \cdots }
\auto
$$
where the remainder of the vertex does not contain 
the anomaly.

The anomaly divergence appears in the limit
$$
\eqalign{&(q_2 + k_1 +k_3)^2  \sim (q_1+q_2+k_1 +k_2)^2 \cr
&\sim (q_1 + k_2  - k_3)^2 \sim q^2 \to 0}
\auto
$$
with $~q_2+k_1 + k_3~(= k_1-q_1+k_3-q_3)$ having a finite light-like
component. Mass-shell constraints resulting from the 
longitudinal integrations, that must be satisfied, are
$$
\eqalign{(k - q_1 + k_1)^2&=(k + q_2 - k_2)^2\cr
& =(k+k_1 -k_2 + k_3)^2=0} 
\auto
$$
All constraints are satisfied, 
with $k=0$, if $q_1^2=q_2^2=k_1^2 =k_2^2$ and
$$
\eqalign{ (q_1 - k_1)& \to -2 l(1,1,0,0) \cr
(q_2- k_2) & \to ~~2 l(1,0,1,0) \cr 
q_3 & \to ~~~l(0,1,-1,0) \cr 
k_3 & \to -l(0,1-2cos {\theta}_{lc}, 1-2 sin {\theta}_{lc}, 0) }
\auto
$$
In the limit $q\to 0$, only the light-like vector 
$k_{lc}=2l (1, cos {\theta}_{lc}, sin {\theta}_{lc}, 0)$
flows through the triangle and the anomaly gives
$$
\Gamma_6 ~~\sim  ~~ {(1 - cos {\theta}_{lc} - sin {\theta}_{lc})^2
~l^2 \over q}  
\auto
$$
It is important that although the mass-shell constraints are 
satisfied in this momentum configuration,
they do not correspond to physical region discontinuities for the 
scattering process of Fig.~7. Instead, as noted above, and discussed in detail
in \cite{arw99}, the corresponding multiple discontinuity is unphysical.   

\section{PROPERTIES OF THE ANOMALY}
 
\subsection{The Space-time Picture}

The momentum configuration described in the previous Section
corresponds to the physical scattering process shown in Fig.~10, which we 
refer to as the ``basic anomaly process''.
\begin{center}
\epsfxsize=2.9in
\epsffile{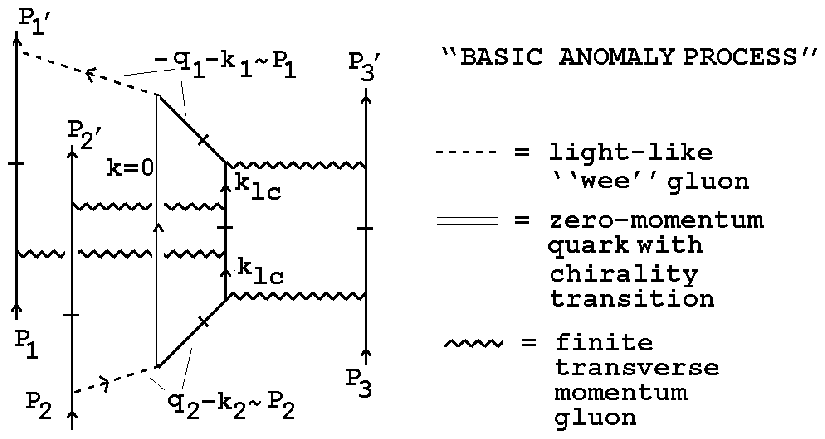}

\vspace{0.1in}

Fig.~10 The physical scattering giving the anomaly.
\end{center}
With the time direction up the page we obtain the following 
space-time picture.
An incoming quark emits a ``wee gluon'' that 
produces a zero-momentum quark together with a
light-like antiquark. The zero-momentum quark undergoes a 
chirality transition while the antiquark's lightlike momentum is rotated to 
$k_{lc}$ by a $t_3$-channel gluon. The antiquark then forward scatters off 
a gluon from the $t_1$ channel before another
$t_3$-channel gluon again rotates the antiquark light-like momentum so that it 
and the zero-momentum quark can 
annihilate
into an 
outgoing wee gluon with spacelike momentum
orthogonal to the initial incoming wee gluon.

We emphasize that the scattering process of Fig.~10, with the
chirality transition, enters the physical region only asymptotically
and then only when the quark mass is zero.

\subsection{Reggeon Ward Identities}

We anticipate that gauge invariance will be manifest via reggeon Ward 
identity cancelations\cite{arw98} 
that could involve the anomaly. For example, the 
diagram of Fig.~11 cancels the anomaly in the diagram 
of Fig.~10 when the ``wee gluons'' are in a color zero state in the 
$t_3$-channel.
\begin{center}
\epsfxsize=1.4in
\epsffile{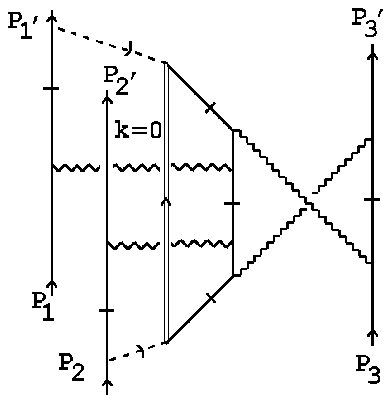}

Fig.~11 A diagram related to Fig.~10 by a reggeon Ward identity
\end{center}
In a color octet state the Ward identity involves
a gluon interaction which can not contain 
a chirality transition. As a result, there is no cancelation
and the reggeon Ward identity is violated. A
non-abelian gauge symmetry is crucial, therefore, for the non-cancelation
of the anomaly.

\subsection{Parity Cancelations}

If an alternative set of quark lines is placed on-shell in the
original maximally non-planar diagram of Fig.~5, a further anomaly 
contribution is obtained. As shown in
Fig.~12, a parity transformation interchanging $P_1$ and $P_2$ relates the
two contributions and hence the anomaly has the opposite sign. 
\begin{center}
\epsfxsize=2.1in
\epsffile{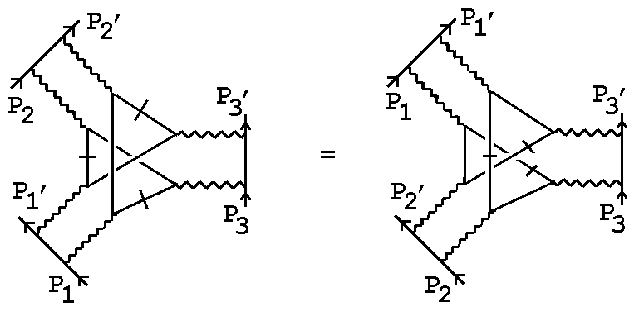}

Fig.~12 Another anomaly contribution with $P_1$ and $P_2$ interchanged.
\end{center}
This does not produce a cancelation in the 
reggeon interaction we have extracted, but does in the 
complete diagram after the transverse 
momentum integrations are performed.

\subsection{When is there no cancelation?}

Because of the discontinuity structure of the amplitudes in which it 
is contained\cite{arw99}, the anomaly vertex conserves signature 
(producing, we anticipate, an even signature pomeron in hadron amplitudes). 
Because of it's parity properties the anomaly 
couples only anomalous color parity (i.e. not equal to the signature)
gluon reggeon combinations. Such exchanges do not 
couple to elementary quark or gluon scattering states and so, in effect,
the anomaly cancels to all orders for such states.
(Note, however, that the cancelation 
occurs after transverse momentum integrations have 
been performed, not in the reggeon interaction.)

With clusters (potentially forming bound states) 
in the initial or final states there can be sufficient structure 
in the couplings to the 
exchanged reggeons that the parity properties of the anomaly do not
produce a cancelation. The anomaly infra-red divergence
will be suppressed by the reggeon
Ward identity zeroes of such couplings, but there will also be 
ultra-violet effects which we do not expect to be suppressed. 
Therefore we anticipate that, in amplitudes 
that contain the anomaly (there must be an even number of anomaly vertices
for color parity to be conserved), 
ultra-violet effects will produce a power violation 
of unitarity bounds by reggeon exchanges. Indeed, we suspect that 
unitarity violation associated with anomaly interactions is 
a core problem for the existence of a bound state S-Matrix in a 
general gauge theory.

\section{THE PROPOSED QCD SOLUTION}

We are currently constructing a solution for QCD based on the properties
of the anomaly discussed above. We can briefly describe 
the essential features that are emerging, as follows.

Using a generalization of the procedure outlined in Section 4, 
a complete set of multi-reggeon scattering amplitudes can be extracted
from high-order particle amplitudes. In superconducting QCD,
with SU(3) color broken to SU(2), all reggeon states with non-zero SU(2)
color have infra-red divergences that exponentiate amplitudes to zero.
If we consider initial scattering reggeon states, with anomalous color parity,
that contain massive gluon reggeons (or quark reggeons) in a reggeon
condensate, we obtain a sub-set of color zero amplitudes that has an overall
logarithmic infra-red divergence.
The divergence occurs when all triple-regge interactions 
that could contain the anomaly, do so. If the divergence is 
factored off the residue is a set of ``physical reggeon amplitudes'' 
in which the condensate appears also in all intermediate and 
final reggeon states (a completeness property for states containing the 
condensate). Ultra-violet effects of the anomaly are not present
in the physical amplitudes.

The reggeon interactions remaining after the 
anomaly divergence is factored off have the general
form shown in Fig.~13.
\begin{center}
\epsfxsize=2.7in
\epsffile{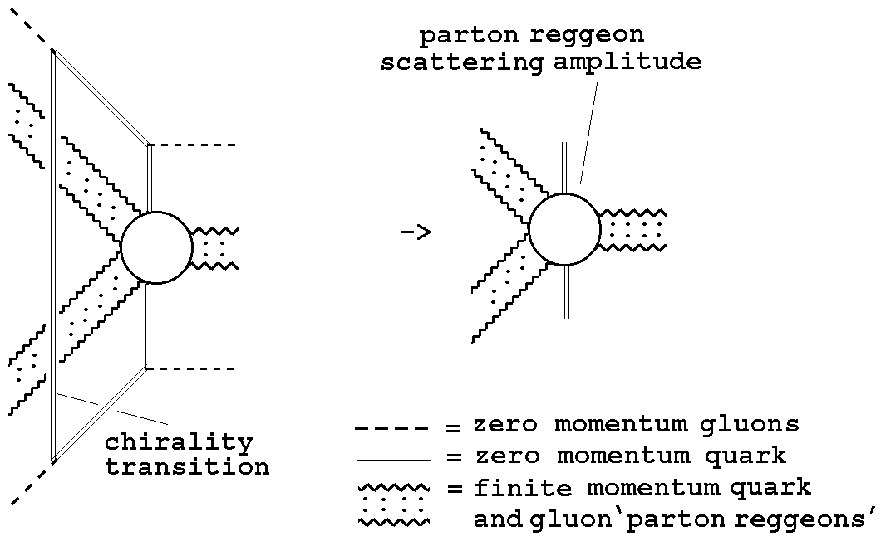}

\vspace{0.1in}

Fig.~13 The ``parton reggeon interaction''.
\end{center}
Comparing with the basic anomaly process of Fig.~10, we note that the 
complete quark loop containing the chirality transition 
carries zero momentum and that the forward scattering
part of the interaction has been replaced by a tranverse momentum 
conserving, ``parton reggeon interaction''.
The  parton scattering lies in the broken part of 
SU(3), while the background anomaly interaction of zero momentum
gluons lies in the unbroken part. These gluons carry the quantum
numbers of the SU(2) winding number current in each $t$-channel.
There is confinement of SU(2) gluons in that the
``parton reggeons'' do not include 
massless gluons - such states are amongst those exponentiated to zero. 

The pomeron is a massive reggeon in the reggeon condensate and 
carries odd color parity as anticipated. 
The BFKL pomeron does not appear. The 
lowest-order triple-pomeron interaction is 
given by diagrams of the form shown in Fig.~14. 
\begin{center}
\epsfxsize=1.5in
\epsffile{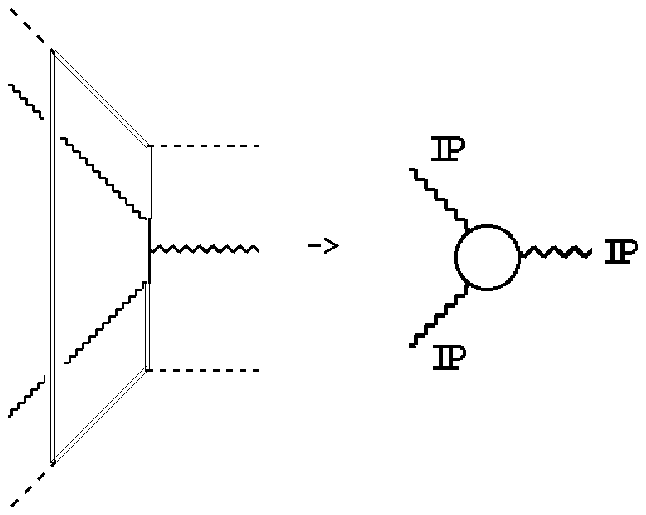}

\vspace{0.1in}

Fig.~14 A diagram contributing to the triple-pomeron interaction.
\end{center}
The reggeon condensate appears to have all the right properties to be 
identified with  
the pomeron condensate of super-critical RFT.

There are massless quark states that 
are the Goldstone bosons corresponding to chiral 
symmetry breaking. Since the unbroken gauge symmetry is SU(2), 
these include\cite{ketal} both pions and ``nucleons''.
The chiral symmetry breaking involves an 
additional chirality violation that is,
in effect, an S-Matrix analogue of the appearance of a chiral condensate.
Additional chirality transitions 
appear in $\pi - \pi$ scattering, within $k_{\perp}$ integrals as 
illustrated in Fig.~15, to compensate
for the helicity-flip of the anomaly interaction.
Although we will not discuss higher-order diagrams here, we hope to
establish a complete correspondence between the reggeon diagrams describing
$\pi - \pi $ scattering in superconducting QCD and the supercritical 
expansion described in Section 3.
\begin{center}
\epsfxsize=2.9in
\epsffile{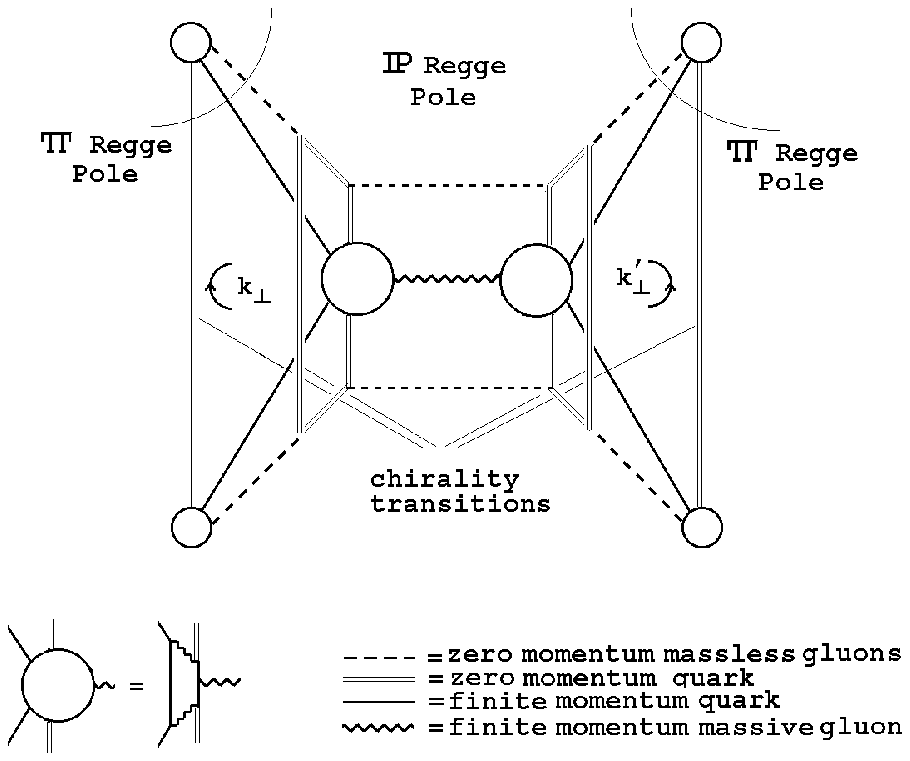}

\vspace{0.1in}

Fig.~15 $\pi - \pi$ scattering - via additional chirality transitions.

\end{center}

As SU(2) gauge symmetry is restored to SU(3) the 
condensate effects should disappear and leave behind 
critical pomeron behavior, as described by RFT. The decoupling of the odd
signature reggeon will then complete the confinement of SU(3) gluons. 
The zero momentum quark and
gluon interactions of Figs.~13 - 15 should be replaced by the interactions 
of a universal wee-parton distribution surrounding a ``hard reggeon 
parton interaction''.
Thus providing the anticipated extension of the parton model.
 
The critical behavior will occur with no
$k_{\perp}$ cut-off
only when the gauge 
symmetry breaking from SU(3) to SU(2) does not destroy the asymptotic freedom
of the theory. This requires extra quarks beyond those observed experimentally.
In fact, the additional quarks required 
could be a color sextet quark sector that is responsible for
electroweak symmetry breaking in the Standard Model. 
However, we will not enlarge on this possibility here.

\mainhead{QUESTIONS}

\noindent M.~LOEWE (P. Universidad Cato'lica, Chile)

\vspace{0.1in}

\noindent Question - Many years ago Bartels 
proposed a picture according to which the wee partons 
obey a diffusion equation. What is the relation of this picture and your 
triple pomeron vertex, including the anomaly?

\vspace{0.1in}

\noindent Answer - Bartels' picture applies to the BFKL pomeron and 
the diffusion is in transverse momentum. Although I discussed the anomaly
within a six-reggeon vertex that could couple three
BFKL pomerons, it cancels in this context. It does appear in my triple-pomeron
vertex, i.e. when the pomeron contains a minimum of four gluons and carries odd
color parity. My pomeron is a regge pole and therefore has the corresponding 
diffusion picture in impact parameter space.

\vspace{0.1in}

\noindent Question - Is there any simple reason why the maximal 
non-planar diagram is the relevant one? 

\vspace{0.1in}

\noindent Answer - It is well-known that non-planar diagrams have the
double spectral function property necessary to produce regge cut couplings in a 
non-gauge theory. For the same reason non-planar diagrams provide the 
essential structure of fermion loop contributions to the
(regge limit) interactions of
photons in QED or gluons in QCD. The planar diagrams, in effect, simply 
regularize divergences appearing in the non-planar contributions. The role of 
the maximally non-planar diagram with respect to the anomaly is, in part, 
an extension of this situation. However, the anomaly also has a kinematic
structure that is intrinsically four-dimensional and (essentially) 
the complexity of 
the maximally non-planar diagram is needed to generate this structure.
 
\vspace{0.1in}

\noindent H.~FRITZSCH (Universit\"{a}t M\"{u}nchen, Germany)

\vspace{0.1in}

Question - There is also the non-perturbative contribution of instantons to 
the anomaly. How would instanton interactions affect your discussion?

\vspace{0.1in}

Answer - The aim of my procedure is to discover 
and regularize the anomaly to produce an S-Matrix  
description of the pomeron and hadron reggeons that 
is unitary in the (multi-)regge region. I should then have 
the complete (multi-regge region) answer, including any contribution 
made by instanton interactions. To explicitly discuss instanton 
interactions it is necessary to start with the euclidean path-integral 
formulation of the theory, continue to Minkowski space, and then (if possible)
extract the regge region 
S-Matrix. It may be, and indeed I expect, that there is a match only
in the special circumstance that we consider SU(3) gauge theory with a 
particular fermion sector.

\end{document}